\documentclass[aps,pra,floatfix,noshowkeys,epsfig,graphics,natbib]{revtex4}%
\usepackage{graphicx}
\usepackage{amsmath}
\usepackage{amsfonts}
\usepackage{amssymb}
\usepackage{comment}

\newcommand{\newc}{\newcommand}
\newc{\beq}    {\begin{equation}}
\newc{\eeq}    {\end{equation}}

\newc{\beqa}    {\begin{eqnarray}}
\newc{\eeqa}    {\end{eqnarray}}
\newc{\bs}    {\section}
\newc{\no}    {\\ \nonumber}

\def\apj{{\em Astrophys. J.  }}

\def\mnras{{ Mon. Not. Roy. Astron. Soc.  }}

\newcommand{\bea}{\begin{eqnarray}}
\newcommand{\eea}{\end{eqnarray}}

\newc{\st}    {\stackrel}

\begin{document}
\title{
Neutrino mass and  ultralight dark matter mass  from the Higgs mechanism}

\author{Jae-Weon Lee}
\email{scikid@jwu.ac.kr}
\affiliation{Department of Electrical and Electronic Engineering, Jungwon University, 85 Munmu-ro, Goesan-eup, Goesan-gun, Chungcheongbuk-do, 28024, Korea.}

\begin{abstract}
We propose a model in which small neutrino masses are generated via Yukawa coupling to a self-interacting  ultralight dark matter (ULDM) field, treated as a pseudo-Nambu-Goldstone boson associated with a heavy Higgs-like field. ULDM has a mass \( m \gtrsim 10^{-22}~\text{eV} \) and a characteristic energy scale \( \tilde{m} \simeq 10~\text{eV} \). The resulting neutrino mass, as well as the mass and self-interaction strength of ULDM, align with cosmological observations. A quantum stability condition for an ULDM effective potential demands a small mass for neutrinos roughly bounded by $\tilde{m}$. The phase transition temperature for the Higgs mechanism can approach the grand unified theory (GUT) scale, potentially inducing the electroweak scale  by reverting the type I seesaw mechanism for Majonara neutrinos. In this framework, neutrino masses can vary with spacetime, a feature that may be experimentally detectable through neutrino oscillation experiments. 
We also explore a scenario in which the tiny ULDM mass arises through radiative corrections via the Coleman-Weinberg mechanism, beginning from a massless field theory. 
 Our model addresses both the neutrino mass and ULDM mass puzzles through a unified approach, providing insights into possible extensions of the Standard Model.
\end{abstract}

\maketitle

\section{Introduction}
Ultralight dark matter (ULDM), often referred to as fuzzy dark matter, ultralight axion, wave-$\psi$  or scalar field dark matter~\cite{1983PhLB..122..221B,Sin:1992bg,Lee:1995af,Matos:1998vk,2000PhRvL..85.1158H,B_hmer_2007}, is gaining attention as a plausible alternative to cold dark matter (CDM), as it retains the advantages of CDM  on scales larger than galaxies while potentially solving the issues associated with cold dark matter at galactic scales
(for a review see ~\cite{Hui:2016ltb,Ferreira_2021,Rindler_Daller_2021,
Lee_2018,Matos_2024}).
The ULDM model has been shown to address many mysteries of galaxies and black holes, such as small-scale issues~\cite{2000PhRvL..85.1158H,Salucci:2002nc,1996ApJ...462..563N,2003MNRAS.340..657D,2003IJMPD..12.1157T}, the typical size and mass of galaxies  ~\cite{Lee:2008ux}, the satellite plane problem~\cite{Park:2022lel}, and the final parsec problem~\cite{koo2024final,Bromley:2023yfi}.
The wave-like nature and a characteristic length scale about $kpc$ of ULDM have been recognized for their ability to address these issues.

The extremely small mass ($m \gtrsim 10^{-22}~eV$) of ULDM can be hypothesized to originate from a  weak breaking of some symmetry~\cite{Sin:1992bg}
or the string axiverse~\cite{Arvanitaki_2010}. For example, in the Coleman-Weinberg mechanism~\cite{CW}, a very small mass can emerge as a classical symmetry is spontaneously broken through quantum corrections.
In Ref. \citealp{Sin:1992bg},
ULDM was proposed as a pseudo-Nambu-Goldstone boson (pNGB).
The self-interactions of ULDM \cite{Lee:1995af}  for a spontaneous symmetry breaking (SSB) could also resolve the tensions with certain observations in the free model (i.e., the fuzzy dark matter model) with $m\simeq 10^{-22}~eV$, such as the Lyman-alpha forest observations~\cite{Irsic:2017yje,Armengaud:2017nkf}.
In \cite{Lee:1995af}, it was proposed that including the self-interaction of ULDM drastically increases the length scale of the model and allows for a wider range of dark matter mass, which can help avoid the tensions of the fuzzy dark matter.

The fact that neutrinos have mass ($m_\nu\lesssim 0.8~eV$) ~\cite{pdg2024} implies the need to extend the Standard Model, as does the existence of dark matter. Therefore, it is natural to develop a model that connects these two mysteries. In particular, the extremely small mass of ULDM  suggests considering a connection with neutrinos, the lightest fermions in the Standard Model. In this sense, our approach in this work can be considered as the simplest extension of the Standard Model, addressing both the masses of neutrinos and ULDM. There are proposals suggesting 
that ULDM can interact with neutrinos ~\cite{Berlin_2016,PhysRevD.97.075017,PhysRevD.97.043001,Davoudiasl_2018}. Our work proposes SSB and a phase transition related to neutrino mass generation by ULDM, constrained by cosmological observations.

In this paper, we propose that the neutrino mass arises from  the Higgs mechanism~\cite{higgs} involving a standard model gauge singlet or a feasibly interacting dark  scalar (ULDM) and its Yukawa coupling to a massless fermion. We assume ULDM to be a pNGB associated with a heavy Higgs-like field.
In the first scenario, ULDM has
mass and self-interaction at a tree level. We also consider another scenario with the Coleman-Weinberg mechanism for SSB, starting with an initially massless dark scalar and a massless neutrino.
In the latter scenario,
the neutrino acquires mass through
SSB, while ULDM gains mass through radiative corrections explaining the small
mass.

In Section II, we discuss a neutrino mass generation scenario via SSB at a tree-level and demonstrate that the mass and self-interaction of ULDM are consistent with cosmological observations. In Section III, we consider another scenario based on the  one-loop Coleman-Weinberg mechanism for the massless scalar with a massless fermion. Finally, in Section IV, we discuss the results.


\section{Self-interacting Ultralight dark matter
with a Broken Symmetry}

The ULDM as a pNGB  can be described by a  real scalar field $\phi$
with an effective action  \cite{Lee:1995af,Chavanis_2011},
\beq
\label{action}
 S=\int \sqrt{-g} d^4x[\frac{-R}{16\pi G}
-\frac{g^{\mu\nu}} {2} \phi_{;\mu}\phi_{;\nu}
 -V(\phi)],
\eeq
where the potential for SSB
can be given by
\beq
\label{Vphi}
V(\phi)=-\frac{m^2 }{2}\phi^2 + \frac{\lambda \phi^4}{4}
\eeq
with  a dimensionless self-coupling constant $\lambda$.
Note the minus sign of the quadratic term.
The evolution of the field is described by the Klein-Gordon equation 
$$
\square \phi+2 \frac{d V}{d\phi^2} \phi=0,
$$
where $\square$ is the d'Alembertian.
It is well known that
after the symmetry is broken, the field attain a
vacuum expectation value 
\beq
\langle\phi\rangle=v\simeq m/\sqrt{\lambda}.
\eeq
One can expand
the potential near $v$ and redefine
the  field around $v$ as $\phi - v\rightarrow \phi$
which has an effective mass $m_{\phi}=\sqrt{2}m$
and an effective coupling constant $\lambda_\phi=\lambda$.
In the Newtonian limit, we can ignore cubic terms in galactic dynamics because odd-power terms like \( \phi^3  \) oscillate rapidly at frequencies on the order of \( m_\phi \) and average out to zero relative to galactic time scales.

From the equation of motion, one can see that the typical field value is approximately \(\phi \sim m_\phi/\sqrt{\lambda}\), and therefore the typical energy density is \(m_\phi^2\phi^2 \sim m_\phi^4/\lambda\).
 It is useful to define the typical energy scale of the scalar field
 as
 \beq
 \tilde{m}\equiv m_\phi/\lambda^{1/4},
 \eeq
which turns out to be of order of $10 ~eV$
from many cosmological constraints (See below of Eq. (\ref{lambdaJ})). 
In the context of thermal field theory, the SSB phase transition occurs at 
\beq
T_c \simeq \alpha v \simeq \alpha m_\phi/\sqrt{\lambda}=\alpha\tilde{m}^2/m_\phi,
\eeq
where the constant $\alpha$ should be $O(1)$, unless there is fine-tuning.
Note that if $\phi$ itself is a Higgs-like field, $T_c$ should be approximately $m_\phi$
not about $m_\phi/\sqrt{\lambda}$. Instead, here we have assumed that $\phi$ is an angular component (similar to an axion) of a Higgs-like field $\Phi$ having SSB
at $T_c$. That is,
\beq
\Phi(x,t) = \left(\langle \Phi \rangle + \Phi_r(x,t)\right)e^{i\phi(x,t)/\langle \Phi \rangle },
\eeq
where $\langle \Phi \rangle$ is the vacuum expectation value of $\Phi$ with a some
appropriate potential,  and $\Phi_r$ is a radial part. 
Since  $\langle\phi\rangle=v=O(|\langle \Phi \rangle|)$ in general, we can expect
$T_c=O(v)$.
Therefore,  ULDM can be a pNGB boson associated with a phase transition of $\Phi$ at $T_c$.
Note that both \(\Phi\) and \(\phi\) undergo their own SSB.
This scenario also explains the small mass $m$ of ULDM as a pNGB. If ULDM is a pNGB, it could be a real field with a soft symmetry breaking potential $V(\phi)$ in Eq. (\ref{Vphi}),
which gives $\langle \phi\rangle=m/\sqrt{\lambda}$
in the $angular$ direction 
of $\Phi$.
In this paper we do not consider a   cosine potential which gives an effective  attractive quartic term ($\lambda<0$).

Many constraints on ULDM model come from the observation of galaxies which is usually nonrelativistic. Thus, it is useful to derive the Newtonian limit of the Klein-Gordon equation.
For this we decompose the real field as ~\cite{Hui:2016ltb}
$$
\phi(t, \mathbf{x})=\frac{1}{\sqrt{2 m_\phi}}\left[e^{-i m_\phi t} \psi(t, \mathbf{x})+e^{i m_\phi  t} \psi^*(t, \mathbf{x})\right],
$$
where $\psi$ satisfies the following  Schr\"{o}dinger-Poisson equation ~\cite{Lee:1995af,B_hmer_2007,Chavanis_2011,Li_2014}
\beqa
\label{spe}
i\hbar \partial_{{t}} {\psi} &=&-\frac{\hbar^2}{2m_\phi} \nabla^2 {\psi} +m_\phi{U} {\psi}+ \frac{\lambda|\psi|^2\psi}{2m_\phi^2}, \no
\nabla^2 {U} &=&{4\pi G} \rho,
\eeqa
where  $U$ is the gravitational potential,
 $\rho=m_\phi |\psi|^2$ is the dark matter density, and we ignore the expansion of the universe. 
 $\psi$ represents a zero-temperature macroscopic wave function for condensed 
ULDM. 
 In this paper we focus on the Thomas-Fermi regime
where we can ignore quantum pressure compared to the pressure from the self-interaction.
 In this limit,   the kinetic term in Eq. (\ref{spe}) can  be ignored, and the exact ground state solution is given by ~\cite{Lee:1995af,B_hmer_2007,Chavanis_2011}
\beq
|\psi|^2=\frac{|\psi(0)|^2 }{\pi r/R_{TF}} sin\left(\frac{\pi r}{R_{TF}}\right),
\eeq
where
the soliton size is 
\beq
R_{TF}
={\sqrt{\frac{\pi\hbar^3 \lambda }{8c G m_\phi^4}}}.
\eeq
This scale represents the typical
size of a core of galaxies.

On the other hand, from the  Schr\"{o}dinger-Poisson  equations one can obtain
 an equation for the
  density contrast $\delta\equiv\delta \rho/\bar{\rho}=\sum_k \delta_k e^{ik\cdot r}$ with a wave vector $k$ and the average matter
  density $\bar{\rho}$ of the universe,
   \beq
  \frac{d^2 \delta_k}{d t^2} +  \left[(c^2_q+c^2_s)k^2-4\pi G \bar{\rho} \right]\delta_k=0,
 \eeq
 where $c_q=\hbar k/2m_\phi$ is a quantum velocity
and   
$c_s = \sqrt{ { \hbar^3 \lambda \bar{\rho} }/{ 2 c m_\phi^4 } }
$ is a sound velocity 
from the self-interaction.
By equating  $c_s^2 k^2$ with $4\pi G\bar{\rho}$
one can obtain the Jeans length~\cite{10.1093/mnras/215.4.575,Chavanis_2011} which 
is the characteristic length over which
the gravity dominates over the repulsion:
\beq
\label{lambdaJ}
\lambda_J = 2\pi / k_J 
={\sqrt{\frac{\pi\hbar^3 \lambda }{2c G m_\phi^4}}}
= 0.978~kpc\left(\frac{\tilde{m}}{10~eV}\right)^{-2}.
\eeq 
Note that $\lambda_J$ is independent of $\bar{\rho}$ and is $2R_{TF}$.
These scales determine the typical size of small galaxies.
From the data of  dSphs  ($R_{TF}\simeq 1kpc$) one can find $\tilde{m}\simeq 7 ~eV$~\cite{Diez_Tejedor_2014}.
There are other cosmological constraints.
To ensure that the oscillation
of ULDM field  behaves as dark matter
at and after the time of matter-radiation equality, 
the quartic term in the potential should be smaller than the energy density at that time. 
This condition leads to  ~\cite{Li_2014,Boudon:2022dxi}
$\tilde{m} \gtrsim 1~eV$.
An observational constraint  associated with cluster mergers provides an upper bound on the dark matter cross-section, 
$\sigma / m_{\phi} \lesssim$ $1 \mathrm{~cm}^2 / \mathrm{g}$  which gives the upper bound \cite{Desjacques_2018}
$ \lambda \lesssim 10^{-12}\left({m_\phi}/{1 {eV}}\right)^{\frac{3}{2}}$.
Considering all these facts~\cite{Arbey_2002}, we can infer that a physical phenomenon at the energy scale of \( \tilde{m} \simeq 10~\text{eV} \) is related to ULDM, and neutrinos are the only known particles that have a similar mass scale. 

\begin{figure}[h]
\includegraphics[width=0.5\textwidth]{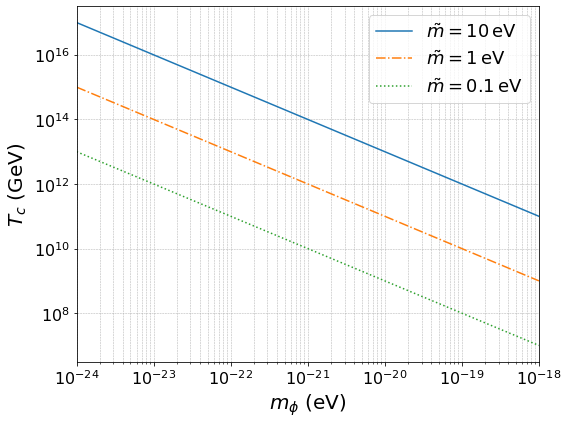}
\caption{ The phase transition temperature $T_c=\alpha v$ for $\alpha=1$, yielding pNGB ULDM $\phi$, is given as a function of $m_\phi$ with various values of $\tilde{m} \equiv m_\phi / \lambda^{1/4}$. }
\label{Tcfig}
\end{figure}

As an example, for the fiducial mass $m_\phi=10^{-22}~eV$ and $\tilde{m}\simeq 10~eV$,
 we obtain $\lambda= 10^{-92}$ and $v= \frac{m_\phi}{\sqrt{\lambda}}= 10^{15}$ GeV.
Interestingly, this energy scale is similar to that of GUT.
In this case
the phase transition temperature
for SSB also becomes $T_c\simeq\alpha v \simeq  10^{15} \alpha ~ GeV$.
More generally, $T_c$ is given by
\beq
T_c\simeq  10^{15} \alpha~ GeV
\left( \frac{m_\phi}{10^{-22}~eV}\right)\left(\frac{\lambda}{10^{-92}}\right)^{-1/2}
=  10^{15} \alpha~ GeV
\left( \frac{10^{-22}~eV}{m_\phi}\right),
\eeq
where we have used the relation $\tilde{m}\simeq 10~eV$ in the last equation. Therefore,
from the observations of galaxies,
the self-interacting ULDM theory reproduces the GUT scale naturally.
This temperature is much higher than the typical Bose-Einstein 
transition temperature for fuzzy dark matter~\cite{Ure_a_L_pez_2009}. 
The GUT scale field value of ULDM can also explain the dark matter density of the present universe ~\cite{Hui:2016ltb}.
For a more massive $m_\phi$, $T_c$ can be lower.
In that case a Pecci-Quinn phase transition
with $T_c \lesssim 10^{13}~GeV$ can be another possible phase transition~\cite{2017PhRvD..95d3541H}. (See Fig. 1.)

We will show how a scalar field in a \(\phi^4\) theory with an energy scale \(\tilde{m} \simeq 10~\text{eV}\) and a very small interaction constant \(\lambda \simeq 10^{-92}\) can be related to neutrino masses when the field \(\phi\) acquires a large vacuum expectation value \(v\) around \(10^{15}~\text{GeV}\).
The Yukawa interaction in this context is given by:
\beq
\mathcal{L}_{\text{Yukawa}} = -y \phi \bar{\nu^c} \nu
\eeq
where \(y\) is the Yukawa coupling constant between the neutrino field $\nu$ and the scalar field.
Since $\phi$ is a real field,
$\nu$ should be a Majonara neutrino.
After SSB the neutrino mass \(m_\nu=yv\)  becomes
\beq
m_\nu = 0.1~eV
\left( \frac{y}{10^{-25}}\right)\left(\frac{v}{10^{15} \mathrm{GeV}}\right),
\eeq 
which implies the SSB giving mass to neutrionos can happen during the GUT phase transition, possibly with the inflation.
In that case ULDM can be
related to the GUT Higgs field $\Phi$. 
 One can obtain the Yukawa coupling $y$ as:
\beq
y = \frac{m_\nu \sqrt{\lambda}}{m_\phi} =  10^{-25}
\left( \frac{m_\nu}{0.1~eV}\right)
\left( \frac{m_\phi}{10^{-22}eV}\right)^{-1}\left(\frac{\lambda}{10^{-92}}\right)^{1/2}
\simeq   
10^{-25}
\left( \frac{m_\nu}{0.1~eV}\right)
\left( \frac{m_\phi}{10^{-22}eV}\right)
\label{y}
\eeq
 for $\tilde{m}\simeq 10~eV$.
Note that $y\simeq m_\nu \lambda^{1/4}/(10~eV)$
and the one-loop quantum correction from the Yukawa
term is $O(y^4)\simeq O(10^{-8}) \lambda \ll \lambda$, which does not
destabilize the potential.
Since the correction is negative, a repulsive self-interaction coupling ($\lambda>0$) at the tree level is required for
stability.
Conversely, if we require the one-loop correction from the Yukawa term \((O(y^4))\) to be smaller than \(\lambda\), we obtain  
\[
m_\nu = y \frac{m_\phi}{\sqrt{\lambda}} \lesssim \lambda^{1/4} \frac{m_\phi}{\sqrt{\lambda}} = \tilde{m}.
\]  
Thus, the stability against quantum corrections demands a small neutrino mass,  
\(m_\nu \lesssim O(10)~\text{eV}\), which is bounded by the energy scale
 $\tilde{m}$ of ULDM.
Another
constraint comes from the propagation of astrophysical neutrinos from distant sources
~\cite{PhysRevD.97.043001},
which is 
\beq
y \lesssim
1.3\times 10^{-11}
\left( \frac{m_\nu}{0.1~eV}\right)^{1/2}
\left( \frac{m_\phi}{10^{-22}eV}\right)^{3/4}.
\eeq
We see that the mass and self-interaction of ULDM are consistent with cosmological observations and safely give
neutrino mass scales observed
with an appropriate $y$.

Another interesting observation is that a combination of $T_c$
and $\tilde{m}$ gives an
 energy scale (for $\alpha=1$)
\beq
\label{TEW}
T_{EW}\equiv \sqrt{\tilde{m} T_{c}}
\simeq 3.1\times 10^{3} GeV
\left( \frac{m_\phi}{10^{-22}~eV}\right)\left(\frac{\lambda}{10^{-92}}\right)^{-3/8},
\eeq
which is similar to the
electroweak scale.This implies that the phase transition of ULDM may provide a hint to resolving the hierarchy problem as well. For a specific example,
one can consider the following Lagrangian in a mass basis
\beq
\mathcal{L}_{\text{Yukawa}} = - \frac{1}{2} \overline{\nu_1^c} yv \nu_1 - \frac{1}{2} \overline{\nu_2^c} M_R \nu_2 + \text{h.c.},
\eeq
where $M_R= O(T_c)$ is a GUT scale mass for a heavy neutrino.  
By reverting the logic of the
 type I seesaw mechanism
with heavy right handed
neutrinos $\nu_R$, 
one can approximately transform the Lagrangian in a flavor basis as
\beq
\mathcal{L}_{\text{Yukawa}} \simeq - \frac{1}{2} \overline{\nu_L^c} yv \nu_L - \overline{\nu_L} \tilde{T}_{EW}\nu_R - \frac{1}{2} \overline{\nu_R^c} M_R \nu_R + \text{h.c.},
\eeq
where $\tilde{T}_{EW}\equiv \sqrt{yvM_R}\lesssim \sqrt{\tilde{m}T_c}$.
Unlike the usual type I seesaw scenario, the left-handed neutrino acquires a mass long before the electroweak phase transition and can induce a new scale \(\tilde{T}_{EW}=O(TeV)\) after the GUT-scale phase transition in the thermal history of the universe
at $T_c\simeq 10^{15}GeV$. 

\section{Coleman-Weinberg Mechanism}
In this section we investigate another example
with the one-loop Coleman-Weinberg mechanism
for generating the mass of ULDM.
The smallness of the ULDM mass and neutrino mass implies a slight breaking of some symmetry
due to a quantum correction.
We start with a Lagrangian that includes a massless fermion $\psi$ and a massless real scalar field $\phi$ 
with a quartic interaction:
\beq
\mathcal{L} =  \frac{1}{2} (\partial_\mu \phi)(\partial^\mu \phi) - V_0(\phi) +i \overline{\psi}  \not{\partial} \psi - y \overline{\psi^c} \phi \psi + h.c.,
\eeq
where the tree level potential $V_0(\phi)=\lambda \phi^4/4$
and we ignore the gravity for simplicity.
At the one-loop level, quantum corrections generate an effective potential for the scalar field.
Using the dimensional regularization, the effective potential $V_\text{eff}(\phi)$ can be computed as ~\cite{CW}:
\beq
V_\text{eff}(\phi)
=
\frac{\lambda}{4}\phi^4
+\frac{3\lambda^2 \phi^4}{64\pi^2} \left( \log \frac{3\lambda \phi^2}{\mu^2}- \frac{3}{2} \right)
- \frac{y^4 \phi^4}{16\pi^2} \left( \log \frac{y^2 \phi^2}{\mu^2} - \frac{3}{2} \right)
,
\eeq
where $\mu$ is a renormalization  scale. 
Note that the fermion contribution is negative. 
From the radiative correction the scalar field can acquire a nonzero vacuum expectation value, \(v\), which minimizes $V_\text{eff}$, i.e.,
\beq
\left.\frac{dV_\text{eff}}{d\phi}\right|_{\phi=v} 
=v^3 \left[ \lambda + \frac{3 \lambda^2}{16 \pi^2} \left( \ln \frac{3 \lambda v^2}{\mu^2} - 1 \right) - \frac{y^4}{4 \pi^2} \left( \ln \frac{y^2 v^2}{\mu^2} - 1 \right) \right] = 0.
\eeq
Solving for $\lambda$ in terms of the other terms and substituting it into $V_\text{eff}$ simplifies the expression
\beq
V_\text{eff}=\left[
\frac{3\lambda^2}{64\pi^2}
- \frac{y^4} {16\pi^2}\right]
\left(ln(\phi^2/v^2)-\frac{1}{2}\right) \phi^4,
\eeq
which spontaneously breaks the symmetry of $V_0$.
To stabilize the 
potential $V_\text{eff}(\phi)$
we require $\lambda^2 \gg y^4$, which  turns out below to be satisfied for the neutrino mass range observed. 
 
The effective mass of the scalar field \(m_\phi\) 
near the true vacuum is given by the second derivative of the effective potential at \(\phi = v\):
\beq
m_\phi^2 \equiv \left. \frac{d^2 V_\text{eff}}{d\phi^2} \right|_{\phi = v}
=
\frac{\left(3\, \lambda^2 - 4\, y^4\right) v^2 }{8\, \pi^2}.
\eeq
The scalar self-interaction coupling constant \( \lambda_\phi \) can be extracted from the fourth derivative of \( V_{\text{eff}}(\phi) \). At the leading order, it is 
\beq
\lambda_\phi \equiv \frac{1}{6} \left. \frac{d^4 V_\text{eff}}{d\phi^4} \right|_{\phi = v}
=  \frac{11 \left(3\, \lambda^2 - 4\, y^4\right)}{48\, \pi^2}
=\frac{11}{6}\frac{m^2_\phi}{v^2}
\simeq 0.069 \lambda^2
\label{lambda2}
\eeq
which is proportional to $\lambda^2$ when $\lambda^2\gg y^4$. Compared to the quartic term,
higher order terms are negligible 
as far as $|\phi-v|\ll v$.

The stability condition $\lambda_\phi>0$
leads to $\lambda^2\gg y^4$.
Unless there is a miraculous cancellation,
$y$ should be much smaller than $\lambda_\phi^{1/4}\gtrsim 10^{-23}$,
which is satisfied for the neutrino mass
$m_\nu=y v < 1 eV$
for $v\simeq 10^{15}GeV$
(see Eq. (\ref{y})).
The effective energy scale $\tilde{m}$ now becomes
\beq
\frac{m_\phi}{\lambda_\phi^{1/4}}=
\left(\frac{3  \left(3 \lambda^2 - 4 y^4\right)}{44 \pi^2}\right)^{1/4} v 
\simeq 0.38\lambda^{1/2}v\simeq 10~eV,
\eeq
which gives a bound for the parameters $\lambda,v$ and $y$
in turn.
For example, from Eq. (\ref{lambda2}), the stability condition  $\lambda_\phi=(m_\phi/\tilde{m})^{4} > 0$ leads to 
\beq
y \lesssim 3.3 \times 10^{-23}
\left( \frac{\tilde{m}}{10~eV}\right)^{-1}
\left( \frac{m_\phi}{10^{-22}eV}\right),
\eeq
which gives a
new bound for neutrino masses $m_\nu \lesssim 33~eV$ for $v\simeq 10^{15}GeV$.
This mass bound is also satisfied for light sterile neutrinos~\cite{Feng:2021ipq}.

\section{Discussions}
We proposed a model in which the masses of neutrinos are generated through the Higgs mechanism by pNGB ULDM.
Since the one-loop correction of the Yukawa coupling induces a negative quartic coupling, a positive self-interaction coupling
of ULDM is inevitable.
We showed that ultralight mass
of dark matter with the GUT scale phase transition temperature can be related to neutrino mass ($\sim 0.1~eV$) and the electroweak scale.
This consistency is nontrivial, considering the vast difference in energy scales (from $10^{-22}~eV$
to $10^{24} eV$) involved in the calculations.

The challenges of the neutrino mass and the hierarchy problem can be reframed in terms of the small mass and self-interactions of ULDM, thereby reducing the overall number of fine-tuning issues in cosmology and particle physics. If ULDM is a pNGB, the small mass of ULDM particles arises naturally.
Since \( m_\phi \) is larger than the current Hubble parameter, there is a possibility that the ULDM field oscillates with a frequency of approximately \( m_\phi \). The oscillation of ULDM
could be described by a plane wave $
\phi\simeq \sqrt{\frac{2 \rho_{\odot}}{m_{\phi}}} \cos \left[ m_{\phi} ( t-\textbf{v}\cdot x)  \right]
$, where $\rho_{\odot}$
is the dark matter density near the sun \cite{Berlin_2016} and $\textbf{v}$ is the velocity of the wave.
Future data from NANOgrav ~\cite{NANOGrav:2023hfp,Chowdhury:2023xvy} 
or LISA ~\cite{amaro2017laser} will possibly confirm this oscillation with gravitational wave detectors.
The oscillation of the field   could induce variations in neutrino masses, which may be detectable in astrophysical or ground based neutrino oscillation experiments
\cite{Berlin_2016,PhysRevD.97.075017,PhysRevD.97.043001} in the near future with an effective Langrangian
$\mathcal{L}_{\text{eff}} = -m_{\nu} \left( 1 +  \frac{\phi}{v} \right) \bar{\nu^c}\nu.$
 Our model addresses both the neutrino and ULDM masses in a unified framework, which may offer insights into possible extensions of the Standard Model.

\subsection{acknowledgments}
The author is thankful to
Sin Kyu Kang for helpful comments.


\end{document}